\documentclass[review]{elsarticle}

\usepackage[utf8]{inputenc}
\usepackage{xparse} 
\usepackage[breaklinks,hidelinks]{hyperref}

\usepackage{amsmath}
\usepackage{natbib}
\usepackage[flushleft]{threeparttable}
\usepackage{pdflscape}
\usepackage{multirow}
\usepackage{lineno}
\usepackage[T1]{fontenc}
\usepackage{textgreek}
\usepackage{booktabs}
\usepackage{graphicx}
\usepackage{siunitx}

\PassOptionsToPackage{hyphens}{url}
\usepackage[hyphens]{url}

\usepackage{physmacros}

\RequirePackage{hyperref} 

\modulolinenumbers[1]

\journal{Astroparticle Physics}

\newiso{Ca}{48}
\newiso{K}{40}
\newiso{Th}{232}
\newiso{Pa}{231}
\newiso{U}{238}
\newiso{Rn}{222}
\newiso{Ra}{226}
\newiso{Pb}{208}
\newiso{Tl}{208}
\newiso{Bi}{214}
\newiso{Ac}{226}
\newiso{Po}{210}
\newiso{Mo}{100}
\newiso{Co}{60}
\newiso{Cf}{252}
\newiso{He}{4}
\newiso{Am}{241}
\newiso{Mg}{24}

\begin{document}

\begin{frontmatter}

\title{Neutron spectrum measurement in the Yemi underground laboratory}

\author[kriss]{Joong Hyun Kim}
\author[kriss]{Sinchul Kang}
\author[kriss]{HyeoungWoo Park}
\author[kriss]{Jungho Kim}
\author[kriss]{Hyeonseo Park\corref{cor1}}
\author[kriss]{Young Soo Yoon\corref{cor2}}
\author[knu]{Hongjoo Kim}
\author[cup]{Yeongduk Kim}
\author[yemi]{Jungho So}
\author[cup]{SungHyun Kim}


\cortext[cor1]{Corresponding author: hyeonseo@kriss.re.kr}
\cortext[cor2]{Corresponding author: ysy@kriss.re.kr}

\address[kriss]{Korea Research Institute of Standards and Science, Daejeon 34113, Republic of Korea}
\address[knu]{Department of Physics, Kyungpook National University, Daegu 41566, Republic of Korea}
\address[cup]{Center for Underground Physics, Institute for Basic Science, Daejeon 34126, Republic of Korea}
\address[yemi]{Yemilab Operation Center, Institute for Basic Science, Gangwon-do 26141, Republic of Korea}

\begin{abstract}
We report on the measurement of neutron energy spectra at the newly established Yemi Underground Laboratory (Yemilab) in the Republic of Korea, designed to host dark matter and rare-event search experiments. 
A high-sensitivity neutron spectrometer was employed, consisting of ten cylindrical \He[3] proportional counters, eight of which were embedded in cylindrical high-density polyethylene moderators of various sizes. 
To quantify and mitigate contributions from internal \alphasym-backgrounds, each detector underwent a dedicated background measurement using a cadmium-shielded box. 
These backgrounds, primarily originating from trace amounts of U and Th in the stainless-steel housings, were characterized and subtracted during data analysis.
Neutron measurements were carried out at three locations within the Yemilab 
between March to October 2023. 
After waveform-based event selection and correction for \alphasym-backgrounds,
neutron count rates were estimated and corresponding energy spectra were reconstructed using the unfolding method.
The total neutron fluence rates were measured ranged from $(3.24 \pm 0.11)$ to $(4.01 \pm 0.10) \times 10^{\minus5}~\fluxunit$,
with thermal and fast neutron components (1 -- 10~\MeV) ranging from $(1.32 \pm 0.05)$ to $(1.51 \pm 0.05) \times 10^{\minus5}~\fluxunit$ and $(0.27 \pm 0.03)$ to $(0.34 \pm 0.10) \times 10^{\minus5}~\fluxunit$, respectively.
\end{abstract}


\end{frontmatter}


\section{Introduction}

Neutrons are a significant source of background in rare-event physics experiments, particularly those conducted in underground laboratories.
In such experiments—including dark matter searches and neutrinoless double beta decay studies, and low-energy neutrino detection, neutron-induced signals can mimic or obscure the rare signals of interest.
Accordingly, accurate characterization of ambient neutron backgrounds is vital for the design and interpretation of these experiments.

In underground settings, the dominant neutron production mechanisms are (\alphasym,\,n) reactions arising from natural radioactivity in the surrounding rock and construction materials, with additional contributions from spontaneous fission and muon-induced spallation. Since neutron fluxes in these environments are extremely low—often several orders of magnitude below surface levels—precise measurements with long exposure times and careful background control are required.

Previously, neutron energy spectra were measured at the YangYang Underground Laboratory in Korea~\cite{PARK2013302} using a Bonner Sphere Spectrometer system based on SP9 proportional counters (Centronic Ltd.) and high-density polyethylene (HDPE) moderators. 
While Bonner Sphere systems are well-established tools for neutron spectrometry~\cite{WIEGEL200236}, their relatively low sensitivity necessitates long integration times to achieve sufficient statistical precision, particularly in deep underground sites.
To address these limitations, we developed a new neutron spectrometer utilizing ten high-efficiency \He[3] proportional counters, each encapsulated with modular HDPE moderators. Compared to the previous system, the new spectrometer provides approximately ten times higher sensitivity, enabling more efficient measurements of low-background neutron fields with reduced statistical and systematic uncertainties.

A crucial aspect of precise neutron measurements is the quantification and correction of internal \alphasym-backgrounds arising from radioactive contaminants within detector construction materials, such as the stainless-steel housings of the counters. We performed dedicated measurements of these internal backgrounds for each detector, employing a cadmium-shielded enclosure to isolate signals associated with internal radioactivity.

Recently, the Yemi Underground Laboratory (Yemilab)—a new deep facility constructed at a depth of 1,000~m (2,500~meter-water-equivalent)—was completed in the Republic of Korea~\cite{Park:2024sio}. We performed neutron spectrum measurements at three distinct locations within the Yemilab from March 20, 2023, to October 18, 2023. This work presents the results of those measurements, including neutron fluence rates and energy spectra, providing essential baseline data for the next generation of rare-event search experiments planned at the Yemilab.

\section{Instrumentation}

\subsection{Neutron Spectrometer with \texorpdfstring{\He[3]}{3He}-Tube proportional counters}

Thermal neutrons are detected in proportional counters via the nuclear capture reaction: 
\begin{equation} 
{}^{3}He + n = \alphasym{} + {}^{3}H + 764 ~keV,
\end{equation}
The reaction products (a 573 keV proton and a 191 keV triton) ionize \He[3] atoms, generating the electron-ion pairs that are collected by the anode wire. When both particles deposit their full energy within the active gas volume, a full-energy peak at 764 keV is observed in the pulse-height spectrum. If one of the particles strikes the detector wall, a wall effect continuum is produced, extending from the full-energy peak down to the energy of the individual particles.

The neutron spectrometer consists of ten cylindrical \He[3] proportional counters 
(Centronic Ltd., Model 50He3/304/25)~\cite{Centronic} paired with eight neutron moderators of varying sizes and materials.
 Each \He[3] tube has a diameter of 2.5~\cm, an active length of 50~\cm, and is filled with \He[3] gas at 4~\atm. The thermal neutron sensitivity measured for these tubes is approximately 81~cps/nv (where 1 nv = 1 neutron/$cm^2$/s), 
 representing nearly an order of magnitude improvement over the spherical SP9 counters (8~cps/nv) used in previous measurements at the YangYang Underground Laboratory~\cite{PARK2013302,Yoon:2021tkv}. 
 The unit cps/nv denotes counts per second per unit neutron flux, where nv represents the product of neutron density (n) and velocity (v).
 Calibration at the Korea Research Institute of Standards and Science (KRISS) thermal neutron facility confirmed consistent detector responses within 1~\percent agreement.

The eight moderators each feature a central bore of 2.55~\cm diameter designed to house the \He[3] tubes. Their dimensions and compositions are summarized in Table~\ref{tab:moderators}. Six moderators are composed purely of HDPE with diameters ranging from 63.5 to 228.6~\mm, suitable for neutron energy coverage from thermal energies up to approximately 200~\MeV. Two composite moderators embed copper layers between HDPE segments to enhance sensitivity to neutrons above 20~\MeV, extending the system’s capability to surface cosmic-ray neutron measurements. 
Specifically, the cylindrical copper shell have thickness of 12.7 mm and 25.4 mm for the CKB170 and CKB180 detectors, respectively.

Two \He[3] detectors operate without moderators (CKB010A and CKB010B), referred to as bare detectors, used primarily for direct thermal neutron flux measurements. To facilitate handling, all moderators—including those with copper layers—are split longitudinally into five sections, with material separations maintained in composite moderators.
To prevent thermal neutron streaming through the assembly interfaces, the moderator sections were fabricated with a stepped, interlocking joint geometry that eliminates direct line-of-sight paths to the detector.

\begin{table}[htbp]
\centering
\caption{Configuration and material composition of the neutron spectrometer moderators. Bare detectors are used without moderators; composite HDPE/Cu/HDPE moderators are designed to enhance high-energy neutron sensitivity.}
\label{tab:moderators}
\begin{tabular}{llccc}
\toprule
Detector No &Detector name & Composition & Diameter (\si{\mm}) & Length (\si{\mm}) \\
\midrule
0\&1 & CKB010A\&B & Bare detector & - & - \\
2 & CKB025 & HDPE & 63.5 & 604 \\
3 & CKB030 & HDPE & 88.9 & 629 \\
4 & CKB040 & HDPE & 102.0 & 642 \\
5 & CKB050 & HDPE & 127.0 & 667 \\
6 & CKB070 & HDPE & 177.8 & 718 \\
7 & CKB090 & HDPE & 228.6 & 769 \\
8 & CKB170 & HDPE/Cu/HDPE & 165.1/114.3/88.9 & 705/654/629 \\
9 & CKB180 & HDPE/Cu/HDPE & 190.5/139.7/88.9 & 731/705/629 \\
\bottomrule
\end{tabular}
\end{table}

Signals from each \He[3] proportional counter are processed through a serise of electronics modules. Each detector is coupled to a charge-sensitive preamplifier (CAEN A1422), which connects to a high-voltage power supply (CAEN DT8033 and DT1470ET) and shaping amplifier (CAEN N1086). The output signals from the shaping amplifiers feed into two 8-channel digitizers with a sampling rate of 250~MS/s (CAEN DT5725S and N6725S). Data acquisition is managed by the CoMPASS software provided by CAEN, which synchronizes signal recording for all detectors, preserves individual parameter setting values including trigger thresholds and timestamps events independently for each channel.

\begin{figure}[htbp]
\centering
\includegraphics[width=0.8\textwidth, trim=0 0 0 5, clip]{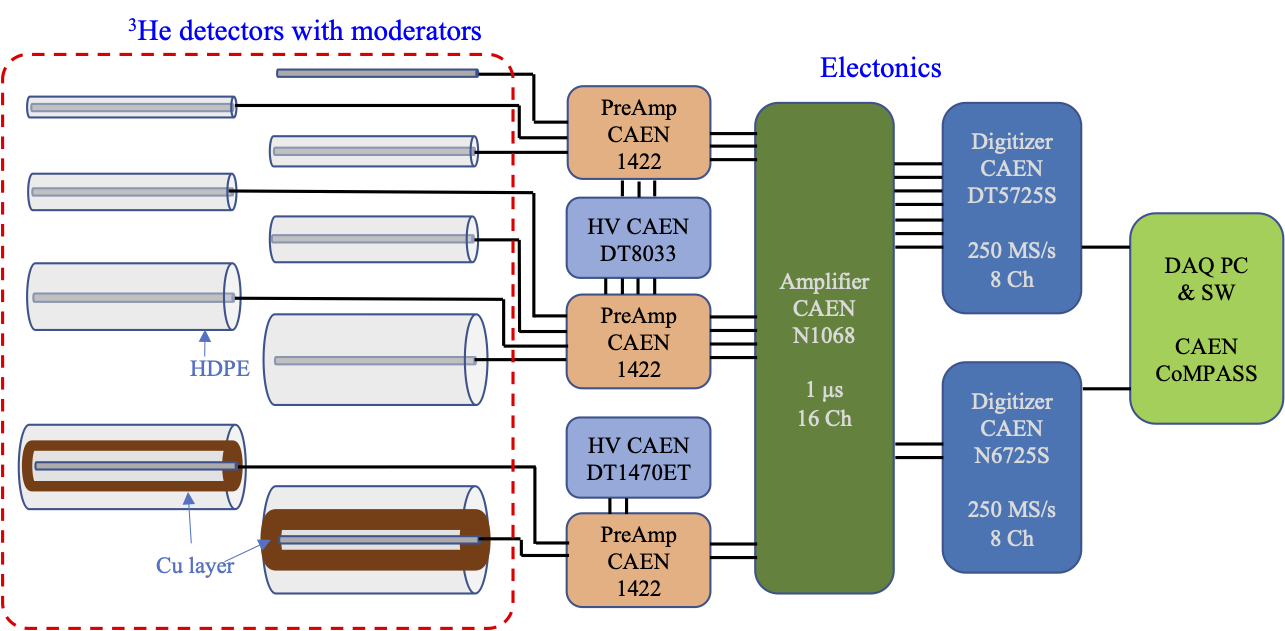}
\caption{Schematic diagram of the detector and electronics system, illustrating the signal path from each \He[3] proportional counter through preamplifiers, shaping amplifiers, digitizers, and data acquisition modules.}
\label{fig:electronics}
\end{figure}

Figure~\ref{fig:electronics} illustrates the detailed signal flow from the \He[3] tubes through preamplifiers and shaping electronics to digitization and data storage.

\subsection{Energy Response Characterization}
\label{sub:energy_response_characterization}

\begin{figure}[htbp]
\centering
\includegraphics[width=0.8\textwidth, trim=0 0 0 5, clip]{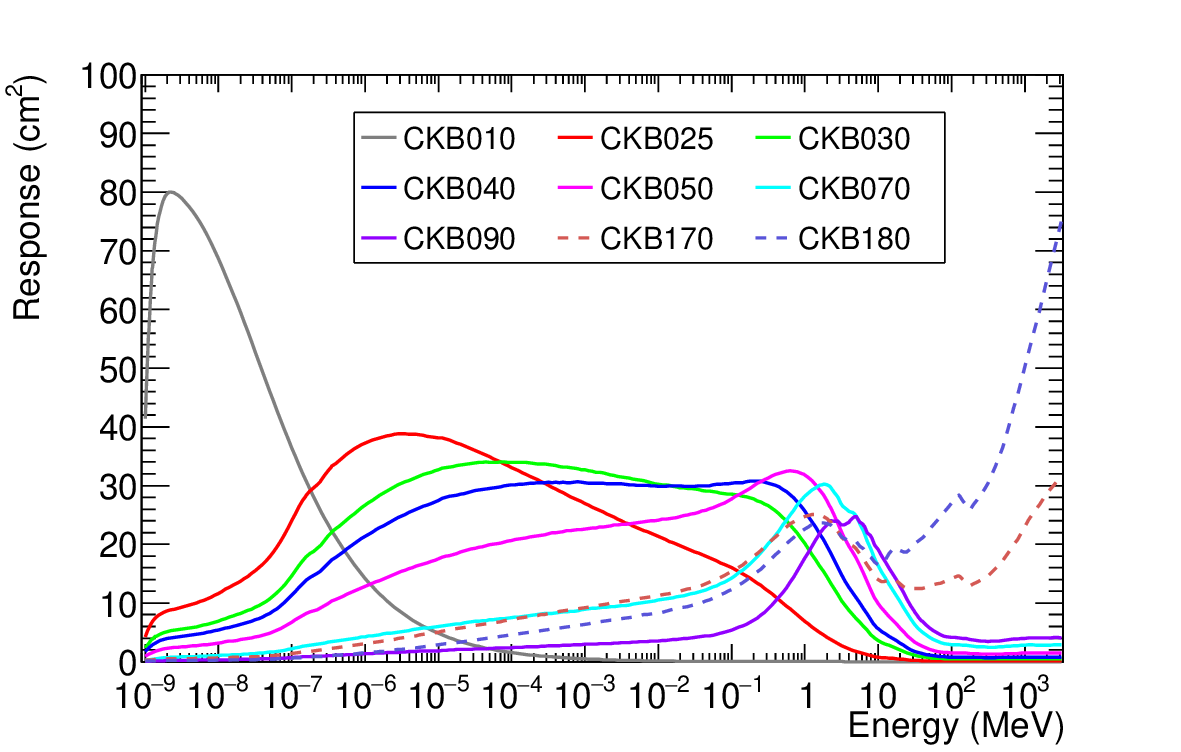}
\caption{Simulated neutron energy response functions for each spectrometer module, calculated using MCNPX. Solid lines represent pure HDPE moderators, while dashed lines denote HDPE/Cu composite moderators with enhanced high-energy sensitivity.}
\label{fig:responsefn}
\end{figure}

Neutron energy response functions of each proportional counter with its moderator were obtained via Monte Carlo simulations using MCNPX version 2.7.0. Figure~\ref{fig:responsefn} presents the energy-dependent response curves for isotropic neutron sources.
The response function acts as the transfer function of the spectrometer, quantifying the combined probability of neutron penetration, thermalization, and subsequent capture. Its shape is determined by the moderator geometry; therefore, employing detectors with distinct response functions enables effective sampling across a wide energy range.

The response of the proportional counters with HDPE moderators covers the energy range from thermal neutrons (20 \meV) up to fast neutrons (approximately 20 \MeV), while the two proportional counters with HDPE moderator embedded with a Cu layer exhibit increasing responses as the neutron energy rises above 20 \MeV.
Calibration of each proportional counter was performed at KRISS using a thermal neutron field and \Am[241]Be neutron sources. 

\section{Measurements}

\subsection{Internal \texorpdfstring{\alphasym}{alpha} Background Measurements}
\label{sec:internal_alpha}

For long-duration neutron measurements in deep underground environments, internal \alphasym-backgrounds—originating from trace uranium and thorium in the stainless-steel housings of \He[3] proportional counters—constitute a notable source of background. Alpha particles from these decay chains can deposit energy within the active gas volume, creating events that mimic neutron captures.

To quantify and correct for these backgrounds, we conducted dedicated measurements at the YangYang Underground Laboratory in 2020. All ten \He[3] counters were placed inside a 1-mm-thick cadmium-shielded enclosure, which effectively suppresses thermal neutrons while allowing detection of signals from internal radioactive decays. Each detector was measured for approximately nine days, totaling 8.9 live days.

\begin{figure}[htbp]
\centering
\includegraphics[width=0.8\textwidth]{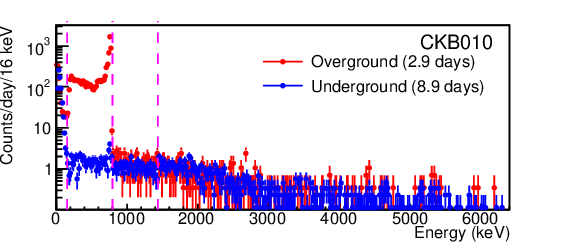}\\
\includegraphics[width=0.8\textwidth]{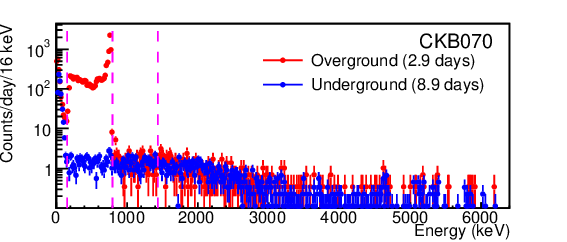}
\caption{Energy spectra measured by two \He[3] detectors, CKB010 (top) and CKB070 (bottom), in underground (blue) and overground (red) locations. The spectra are normalized to daily count rates (counts/day/16keV). Vertical dashed lines indicate the boundaries at 160~keV, 800~keV, and 1.44~MeV, marking analysis regions.
}
\label{fig:alphabgch1}
\end{figure}

\begin{figure}[htbp]
\centering
\includegraphics[width=0.8\textwidth]{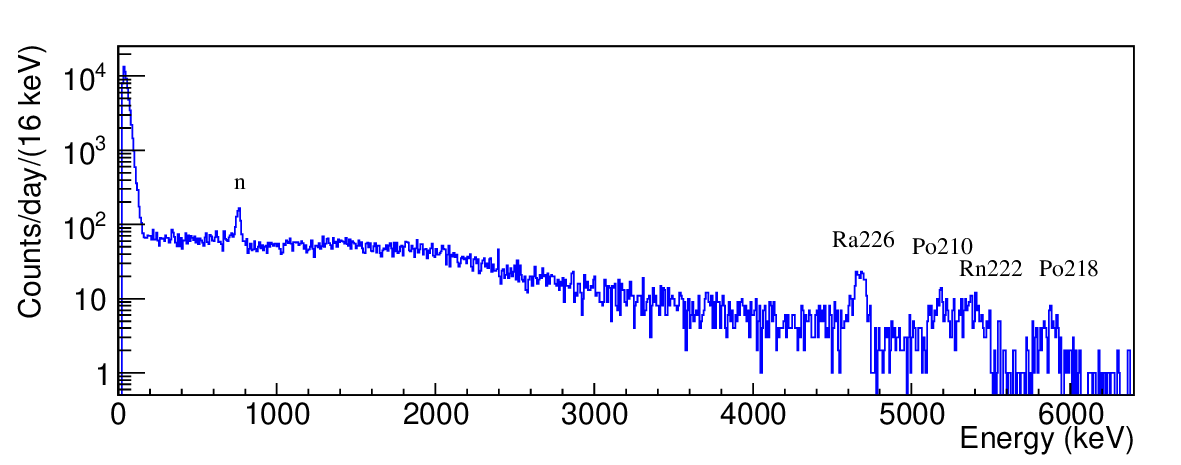}
\caption{Summed energy spectrum from all ten \He[3] detectors during the cadmium-shielded background run. Distinct \alphasym{} peaks from \Ra[226], \Po[210], \Rn[222], and \Po[218] are visible. The small 764 keV peak indicates residual capture of epithermal neutrons penetrating the shield.}
\label{fig:alphabgsum}
\end{figure}

Figure~\ref{fig:alphabgch1} presents example energy spectra for detectors CKB010A and CKB070, comparing underground (blue) and overground (red) data. In underground measurements, a residual 764~keV neutron capture peak is still observable—likely due to neutrons above the cadmium cutoff (approximately 0.5~eV)—while the internal \alphasym-background is seen as a flat continuum above 800~keV and discrete peaks around 4.8–6.0~MeV, corresponding to alpha decays from \Ra[226], \Po[210], \Rn[222], and \Po[218] (see Figure~\ref{fig:alphabgsum}).

The average \alphasym-background rates in the neutron signal region (160–800~keV) were estimated using two independent methods: a constant-rate extrapolation from the high-energy region and a likelihood-based fit combining simulated neutron and background components. Both approaches yielded consistent background rates, which were subsequently used to correct neutron measurements at the Yemilab.

Figure \ref{fig:alphabgsum} identifies the specific \alphasym{}-emitting isotopes identified in the summed spectrum. A residual capture peak at 764 keV is observed; since the 1-mm cadmium layer absorbs neutrons below $\sim$0.5 eV, this feature confirms the presence of ambient epithermal and fast neutrons that penetrated the shield.

\subsection{Neutron Measurements at the Yemilab}

\begin{figure} [!htb]
\begin{center}
\includegraphics[width=0.47\textwidth, trim=0 0 0 5,clip ]{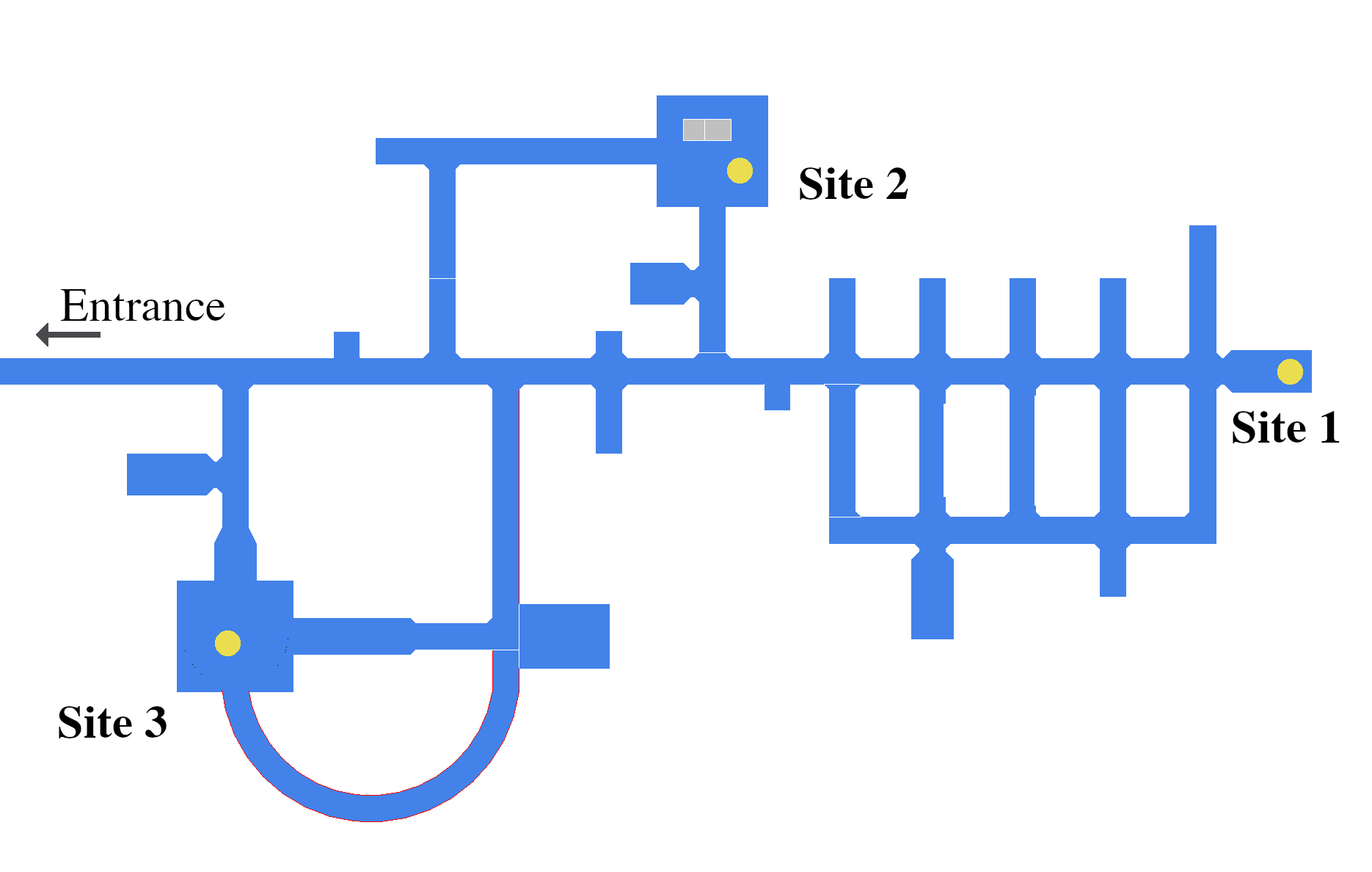}
\includegraphics[width=0.47\textwidth, trim=0 350 0 5,clip ]{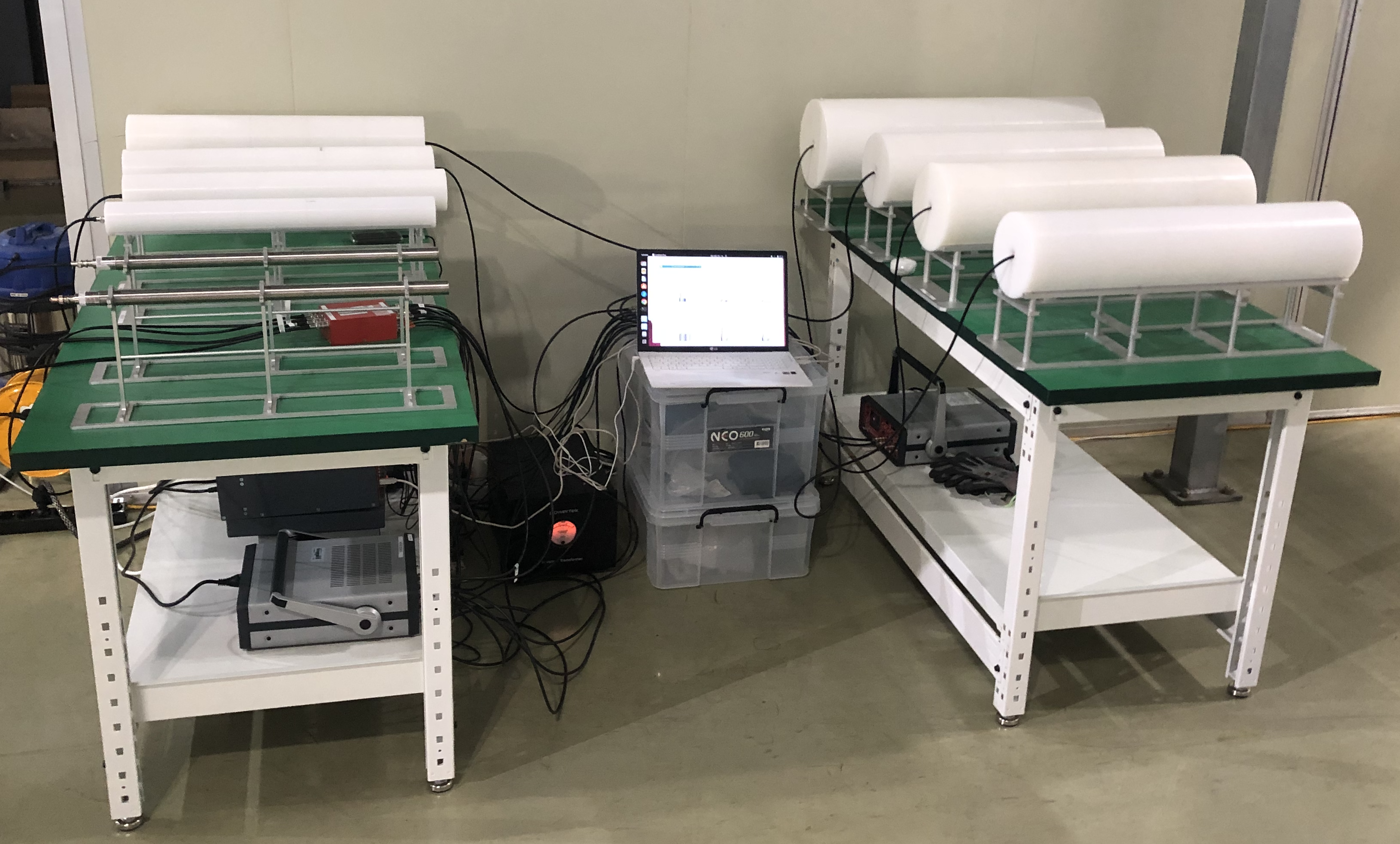}
\caption{(Left) Layout of the Yemilab indicating the three measurement sites (yellow dots). (Right) Experimental setup at Site~2 showing the arranged detector modules.}
\label{fig:setup}
\end{center}
\end{figure}

Neutron spectra were measured at three locations within the Yemilab, labeled Site~1, Site~2, and Site~3 (Figure~\ref{fig:setup}). Site~1 is located near the COSINE experimental area at the tunnel’s far end \cite{2024arXiv240913226C}; Site~2 is in the AMoRE preparation area near the middle of the facility \cite{PhysRevLett.134.082501}; and Site~3 is situated at the base of a large cavity near the tunnel entrance. 
The sites separated by roughly 100~m, with the physical overburden of approximately 994 m, 1,029 m, and 980 m for Sites 1, 2, and 3, respectively. Since the local rock density profile is not yet fully characterized, a nominal depth of 2,500 m.w.e. is adopted for all three locations \cite{Park:2024sio} .

The spectrometer was initially deployed at Site~1, where \Am[241]Be calibration source runs were used to determine the 764~keV full-energy peak and calibrate detector thresholds. After calibration, background neutron data were acquired at all three locations in sequence.
Detector gain stability was monitored by tracking the centroids of the 764 keV neutron capture. Gain drift remained below 1\percent{} throughout the acquisition period, confirming system stability.

At each site, 3–4 data acquisition runs were performed, with individual run durations ranging from 14 to 35 days, depending on operational conditions and access. The total live times were 61.8 days (Site~1), 84.6 days (Site~2), and 49.7 days (Site~3), for an aggregate of 196.1 days. During two runs, data from two detectors were excluded due to hardware issues, resulting in slightly reduced livetimes for those channels. All data were processed on a run-by-run basis, including event selection, \alphasym-background subtraction, and conversion to count rates by dividing the total counts by the active measurement duration.

The following sections detail the event selection procedure, background-corrected count rate estimation, and spectral unfolding analysis leading to the reconstruction of the ambient neutron energy spectrum at each measurement site.

\section{Results and Analysis}

\subsection{Event Selection Using Waveform Analysis}
\label{sec:event_selection}

To efficiently discriminate valid neutron-induced events from electronic noise and spurious signals, waveform-based event selection criteria were applied to digitized pulse traces. Each waveform was characterized by four parameters: 
(1) the pulse width defined as the time difference between the 50\% rise and 50\% decay points ($\Delta T_{\text{DT50--RT50}}$),  
(2) the rise time measured between 10\% and 90\% of the pulse amplitude,  
(3) the decay time measured between 90\% and 10\% of the pulse amplitude, and  
(4) the time at which the pulse reaches maximum amplitude.

Pulse width serves as an effective discriminator as it correlates directly with charge collection time of the ionization electrons. Neutron capture reaction products (protons and tritons) generate distinct ionization signal profile, whereas backgrounds like microphonics typically exhibit much shorter or irregular durations.
We also evaluate other waveform parameters, including rise time (time difference between 90 \percent{} and 10 \percent{} of maximum signal during rising) and decay time (time difference between 10 \percent{} and 90 \percent{} of maximum signal during decaying). 
However, these proved less effective because specific electronics noises mimicked the characteristics of genuine neutron events. Consequently, pulse width provided the most reliable separation between signals and background across all detectors.

\begin{figure}[htb]
\centering
\includegraphics[width=0.47\textwidth, trim=0 0 0 5, clip]{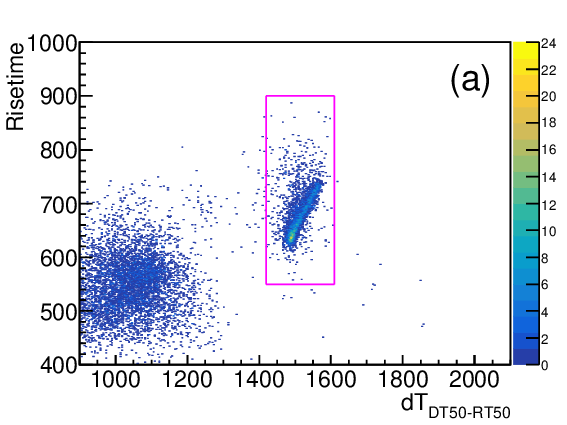}
\includegraphics[width=0.47\textwidth, trim=0 0 0 5, clip]{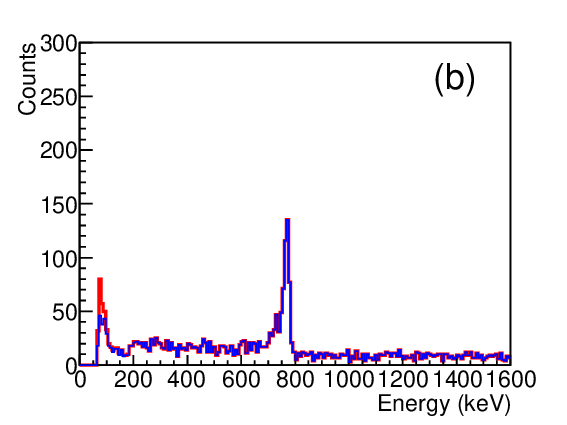}\\
\includegraphics[width=0.47\textwidth, trim=0 0 0 5, clip]{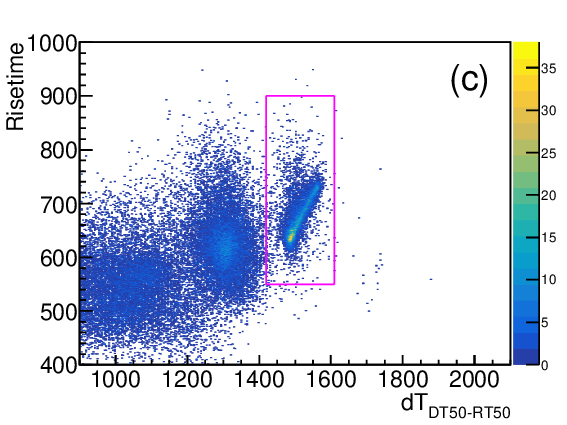}
\includegraphics[width=0.47\textwidth, trim=0 0 0 5, clip]{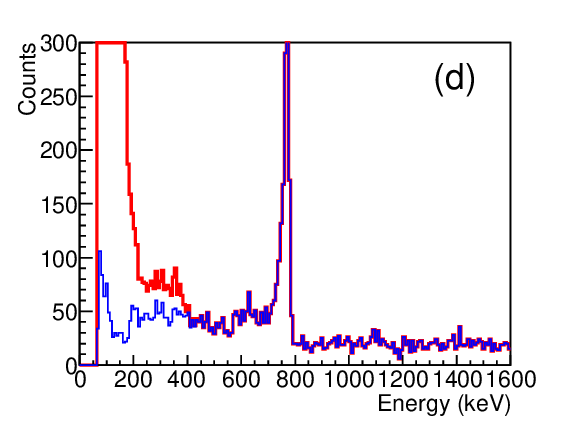}\\
\caption{Waveform-based event selection for two representative datasets: (a, c) Scatter plots of event pulse width $\Delta T_{\text{RT50--DT50}}$ vs. rise time, illustrating neutron-like event clusters within the selection region (magenta box); (b, d) Corresponding energy spectra before (red) and after (blue) event selection, demonstrating noise suppression while preserving neutron signals.}
\label{fig:evtselection}
\end{figure}

Scatter plots of these parameters were used to define selection regions that isolate neutron-like events from noise. Representative examples are shown in Figure~\ref{fig:evtselection}; panel (a) illustrates a clean neutron signal cluster in nominal data, while panel (c) shows data contaminated by noise, forming distinct additional clusters outside the selection box. Corresponding energy spectra before (red) and after (blue) waveform selection are shown in panels (b) and (d). In noise-contaminated data, the selection efficiently removes non-physical events while preserving the neutron signal peak. For clean datasets, event selection produces negligible impact.

Because noise characteristics varied across runs and detectors—occurring intermittently—the event selection criteria were applied individually for each detector and each data run to ensure consistent background suppression while maintaining high neutron efficiency.

\begin{figure}[htb]
\centering
\includegraphics[width=0.9\textwidth, trim=0 0 0 0, clip]{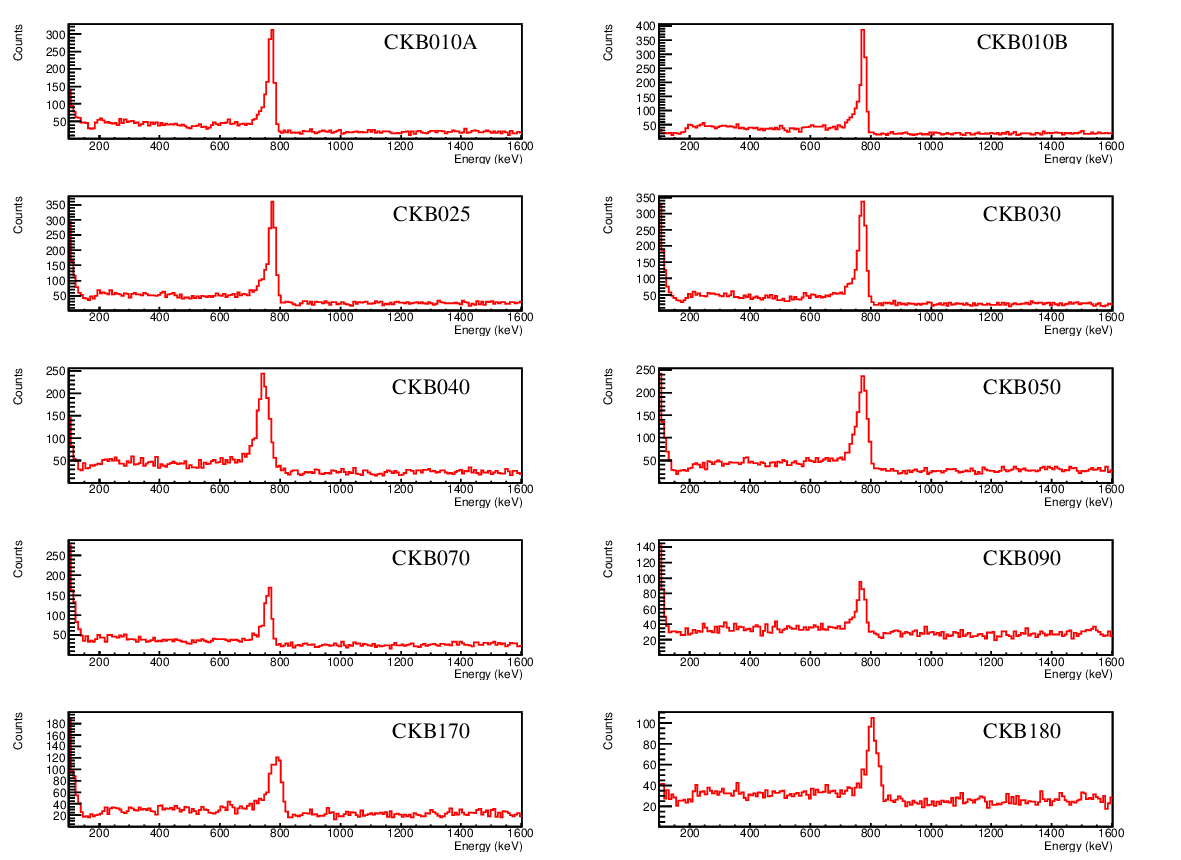}
\caption{Energy-calibrated signal distributions from all detectors at Site~2 accumulated over 82 days of live time. Calibration is based on the 764~keV full-energy peak from \Am[241]Be source runs.}
\label{fig:Site2_counts}
\end{figure}

Figure~\ref{fig:Site2_counts} displays the energy-calibrated spectra accumulated from all detectors at Site~2 over 82 days of live time, calibrated using the 764~keV full-energy peak from \Am[241]Be source runs. The spectra show a relatively flat background above 800~keV attributed to internal \alphasym-background, as discussed in Section~\ref{sec:internal_alpha}.

\subsection{Neutron Count Rate Estimation}

Following event selection, corrections for internal \alphasym-background were applied to the data from each detector and measurement site. Neutron count rates were calculated in the energy window 160–800~keV and computed as a rate over the active measurement duration $T$:
\begin{equation} 
N = \frac{1}{T} \int_{E_n} \bigl(n(E) - Bg\bigr) \, dE,
\end{equation}
where $n(E)$ is the measured energy spectrum and $Bg$ is the estimated background contribution from internal alpha events.

Although anomalous noise below 500 keV appeared in a subset of detectors, a consistent integration window ($E_{\text{sel}}$= 690 -- 800 keV), was applied to the entire array.
The total neutron rate was evaulated using a correction factor, $C_r$, derived from the calibration source spectra, $S(E)$:
\begin{equation} 
C_r = \frac{\int_{E_n} S(E) \, dE}{\int_{E_{\text{sel}}} S(E) \, dE}.
\end{equation}
The corrected neutron count rate, $N_{c}$, for all detectors was then computed as:
\begin{equation} 
N_c = C_r \cdot \frac{1}{T} \int_{E_{\text{sel}}} \bigl(n(E) - Bg\bigr) \, dE.
\end{equation}

\begin{figure}[htbp]
\centering
\includegraphics[width=0.48\textwidth, trim=0 0 0 5, clip]{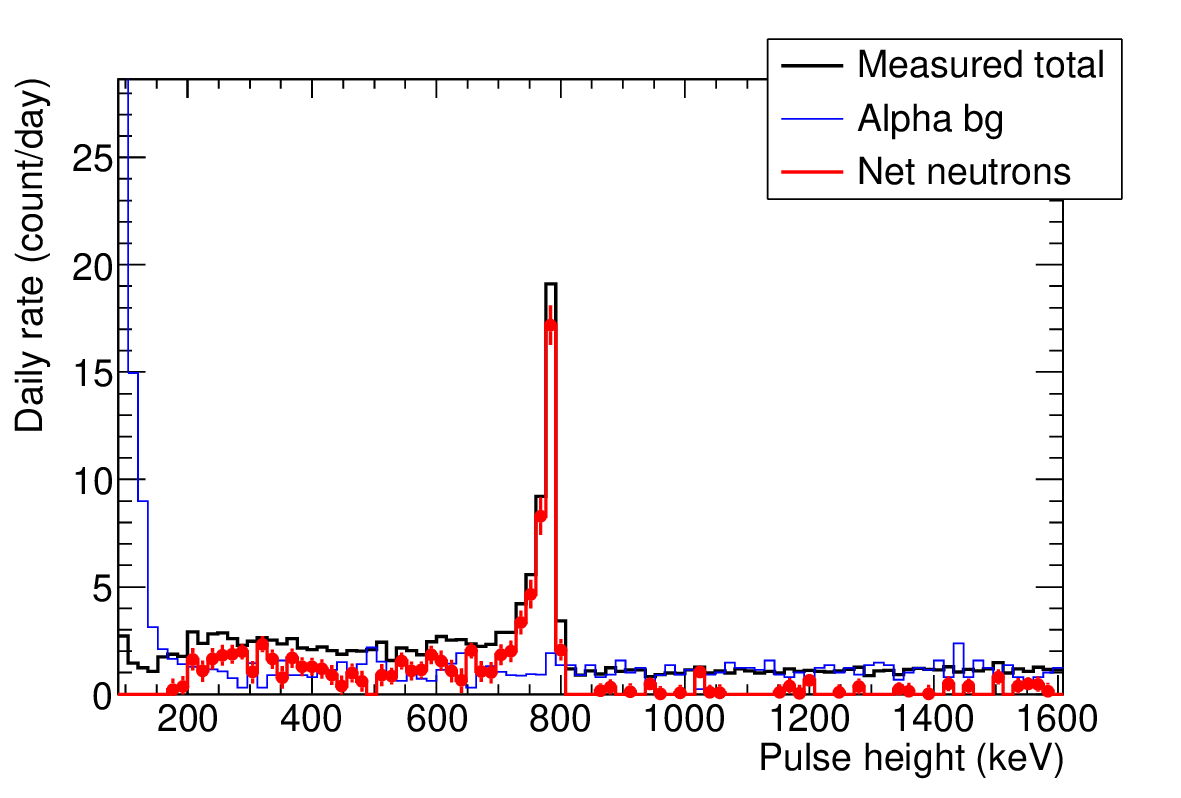}
\includegraphics[width=0.48\textwidth, trim=0 0 0 5, clip]{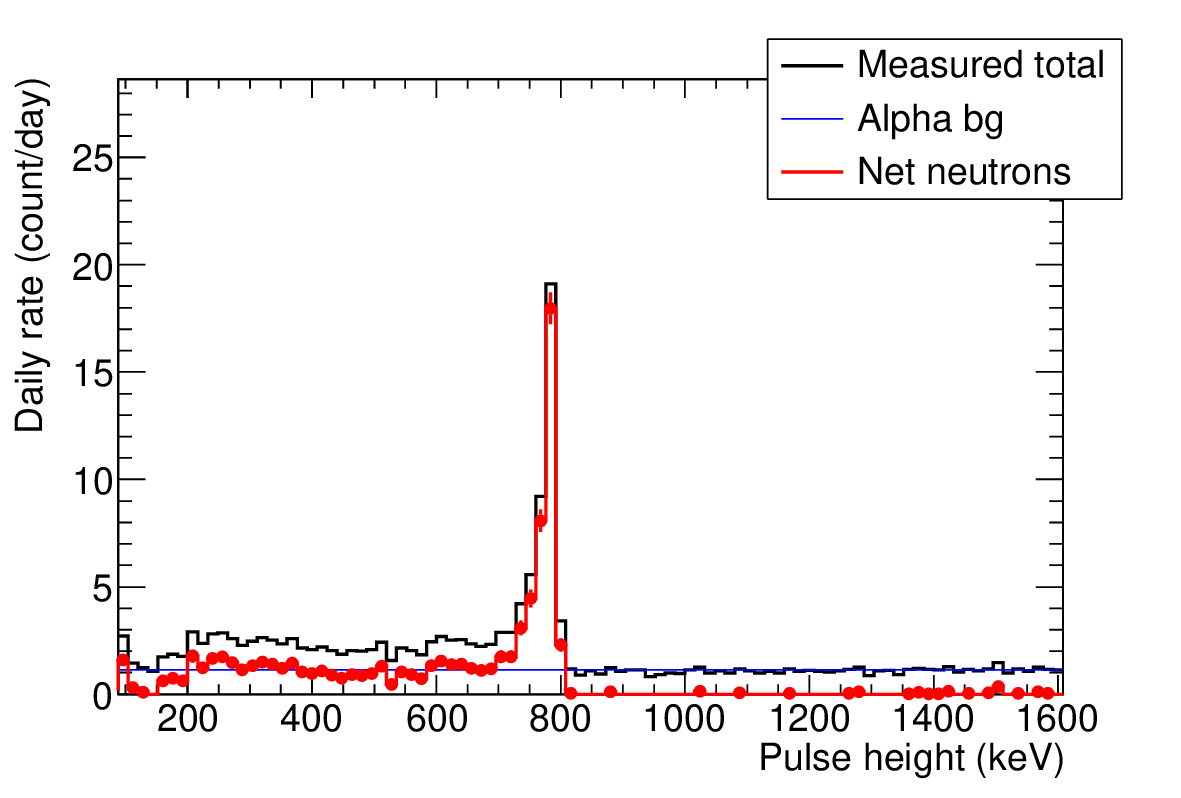}
\caption{Extraction of neutron signals after internal \alphasym-background subtraction at Site~2. Left: subtraction using measured \alphasym-background spectrum. Right: subtraction using average \alphasym rate. Black points show observed data; red lines represent neutron signals accounting for corrections; blue lines indicate background estimates.}
\label{fig:nratefit}
\end{figure}

To validate rate estimation, both the detailed measured \alphasym-background spectrum and an average background rate were used within a likelihood fit framework. Figure~\ref{fig:nratefit} compares results from background subtraction using the measured spectrum (left) and the average background rate (right), demonstrating statistical consistency within 1$\sigma$. For final analyses, the average background was chosen to minimize statistical fluctuations.

Neutron rates were also cross-checked via a binned maximum likelihood fit, consistent with previous Bonner Sphere Spectrometer analyses~\cite{Yoon:2021tkv}. Given the improved counting statistics of the current dataset, direct counting methods were adopted for the final rate evaluation.

\begin{figure}[htbp]
\centering
\includegraphics[width=0.9\textwidth, trim=0 0 0 5, clip]{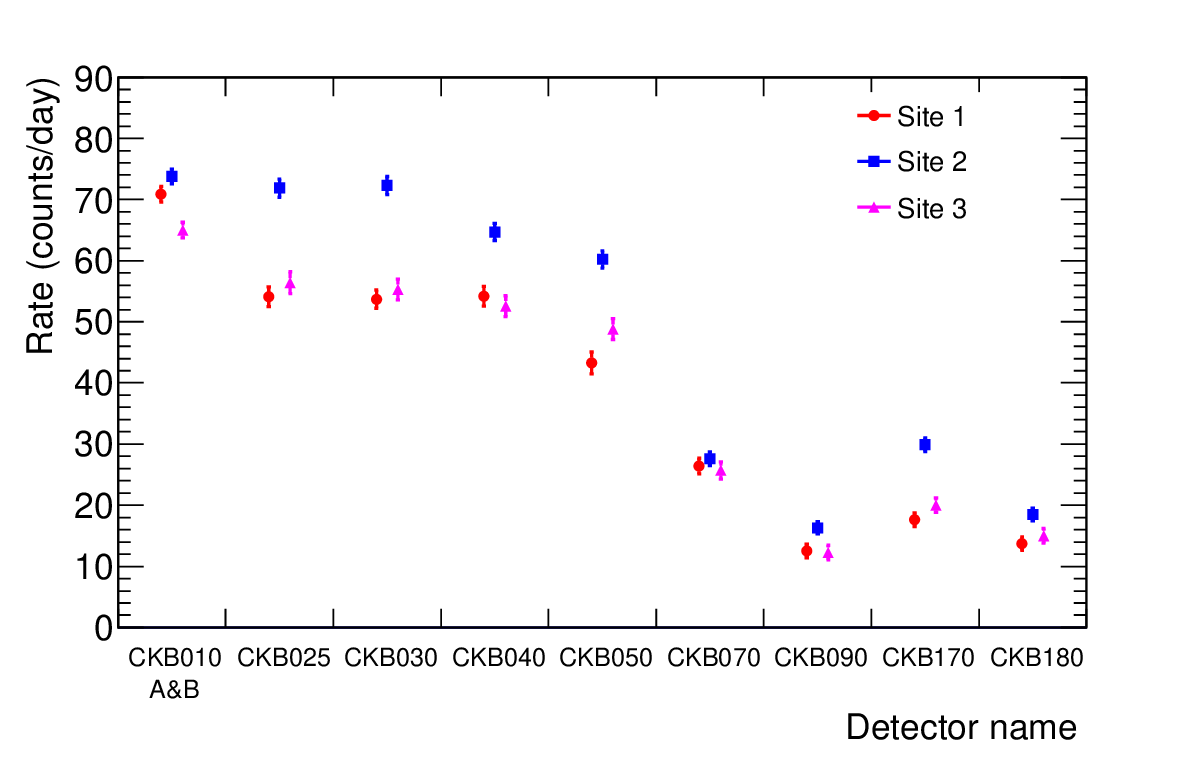}
\caption{Daily neutron count rates measured by each detector and moderator at the three Yemilab sites. 
Values for the two bare detectors (CKB010A and CKB010B) are averaged; individual counts rates from these two units were verified to be mutually consistent within statistical uncertainties at all sites.
Vertical error bars represent statistical uncertainties.}
\label{fig:dailyrate}
\end{figure}

Figure~\ref{fig:dailyrate} presents daily neutron count rates for all detectors across the three measurement sites. Sites 1 and 3 demonstrate stable and comparable neutron rates (within approximately 12\%), while Site 2 shows elevated rates by up to 25\% in certain detectors.

\subsection{Unfolding Analysis}

Neutron energy spectra were reconstructed via an unfolding analysis implemented using the Maximum Entropy method (MAXED) code~\cite{REGINATTO2002242}. The analysis utilized detector response functions derived from MCNPX simulations as detailed in Section~\ref{sub:energy_response_characterization}.
The rising responses of copper-moderated detectors above 20 MeV poses as risk of unphysical high-energy solutions. To mitigate this, we employed MAXED with physical priors typical of underground environments to penalize the divergence tails. Furthermore, we explicitly monitored the optimizer logs to verify that the solution converged to table, physical minima with the required chi-squared limits.

To reduce bias from prior spectrum assumptions, multiple input spectra were tested, including measured spectra from tunnels A5 and A6 at YangYang, ground-level neutron spectra, and composite models combining Maxwellian, 1/E, and fission components. The resulting unfolded spectra were robust against these variations, with the A5 tunnel spectrum adopted for the final results.

\begin{figure}[htbp]
\centering
\includegraphics[width=0.9\textwidth, trim=0 0 0 5, clip]{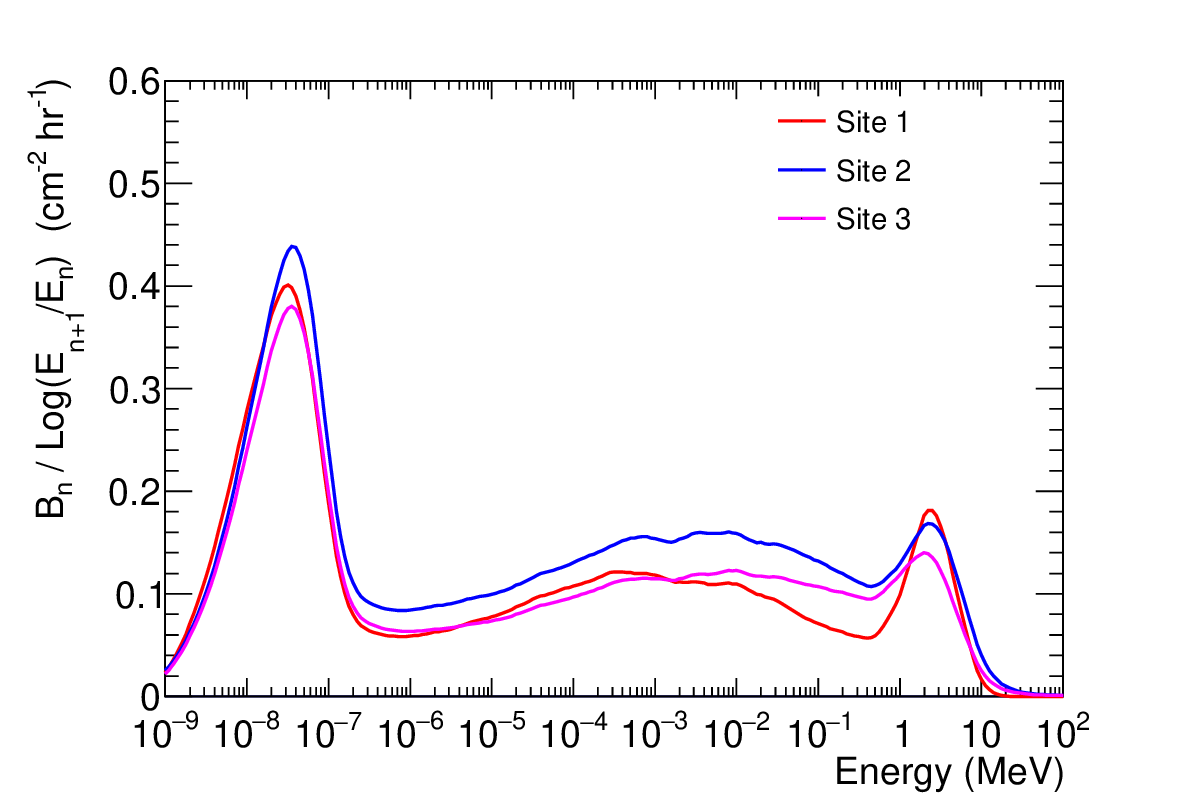}
\caption{Unfolded neutron energy spectra at all three Yemilab sites obtained with the MAXED unfolding method. Distinct features corresponding to thermal, epithermal, and fast neutron populations are visible at each location.}
\label{fig:unfolding}
\end{figure}

Figure~\ref{fig:unfolding} displays the unfolded neutron energy spectra at all three sites, clearly exhibiting thermal peaks, fast neutron components, and broad epithermal regions.

\subsection{Neutron Fluence}

\begin{figure}[htbp]
\centering
\includegraphics[width=0.9\textwidth, trim=0 0 0 5, clip]{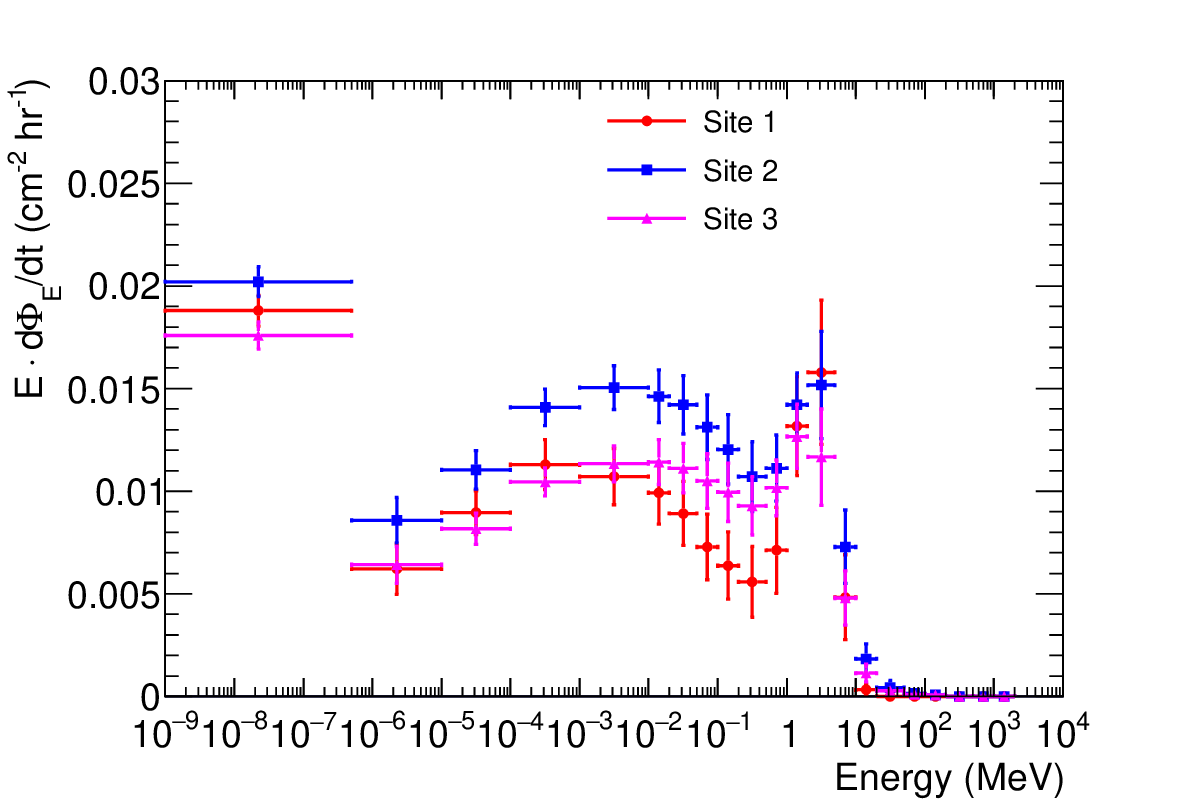}
\caption{Neutron fluence spectra at the three measurement sites in the Yemialb presented in lethargy units, highlighting spectral regions of thermal, epithermal, and fast neutrons.
Vertical error bars represent the statistical uncertainty propagated by the unfolding method.}
\label{fig:YemiFluence}
\end{figure}

Neutron fluence spectra, expressed in lethargy units suitable for logarithmic scale presentation, were computed from the unfolded results (Figure~\ref{fig:YemiFluence}). The spectra exhibit expected thermal ($10^{\minus3}$ -- 0.5~\eV), epithermal (1~\eV -- 0.5~\MeV), and fast (0.5 -- 10 \MeV) neutron components, the last being dominated by (\alphasym, \,n) reaction neutrons.
While spectral shapes are broadly consistent across measurement sites, Site~2 displays a pronounced enhancement in the epithermal region.

Table~\ref{tab:fluence_summary} summarizes integrated neutron fluence rates. While thermal and fast neutron fluences are similar across all sites, Site~2’s total neutron fluence is approximately 25\% higher due to an elevated epithermal flux. 
Systematic uncertainty was estimated by unfolding with three prior spectra: the reference spectrum from the YangYang underground laboratory, a synthetic model combining a Maxwellian thermal peak with a flat spectrum up to 1 \GeV, and a $D_{2}O$-moderated fission spectrum. 
The unfolding results showed systematic uncertainties of 3\percent{} for the thermal neutron fluence and approximately 8\percent{} for the fast neutron component in 0.5--10 \MeV energy range.

\begin{table}[htbp]
\centering
\caption{Integrated neutron fluence rates ($\times10^{\minus6}\,\mathrm{cm^{\minus2}\,s^{\minus1}}$) for thermal, fast, and total neutron components at the three measurement sites in the Yemilab. Statistical and systematic uncertainties are included.}
\label{tab:fluence_summary}
\begin{tabular}{lccc}
\toprule
Site & Thermal ($10^{\minus3}$--0.5~eV) & Fast (0.5--10~MeV) & Total \\
\midrule
Site 1  & 14.1 \pmsym 0.6 & 3.2 \pmsym 0.5 & 32.4 \pmsym 1.1 \\
Site 2  & 15.1 \pmsym 0.5 & 3.4 \pmsym 0.4 & 40.1 \pmsym 1.0 \\
Site 3  & 13.2 \pmsym 0.5 & 2.7 \pmsym 0.3 & 32.5 \pmsym 0.8 \\
\bottomrule
\end{tabular}
\end{table}

\begin{figure}[htbp]
\centering
\includegraphics[width=0.8\textwidth, trim=0 0 0 5, clip]{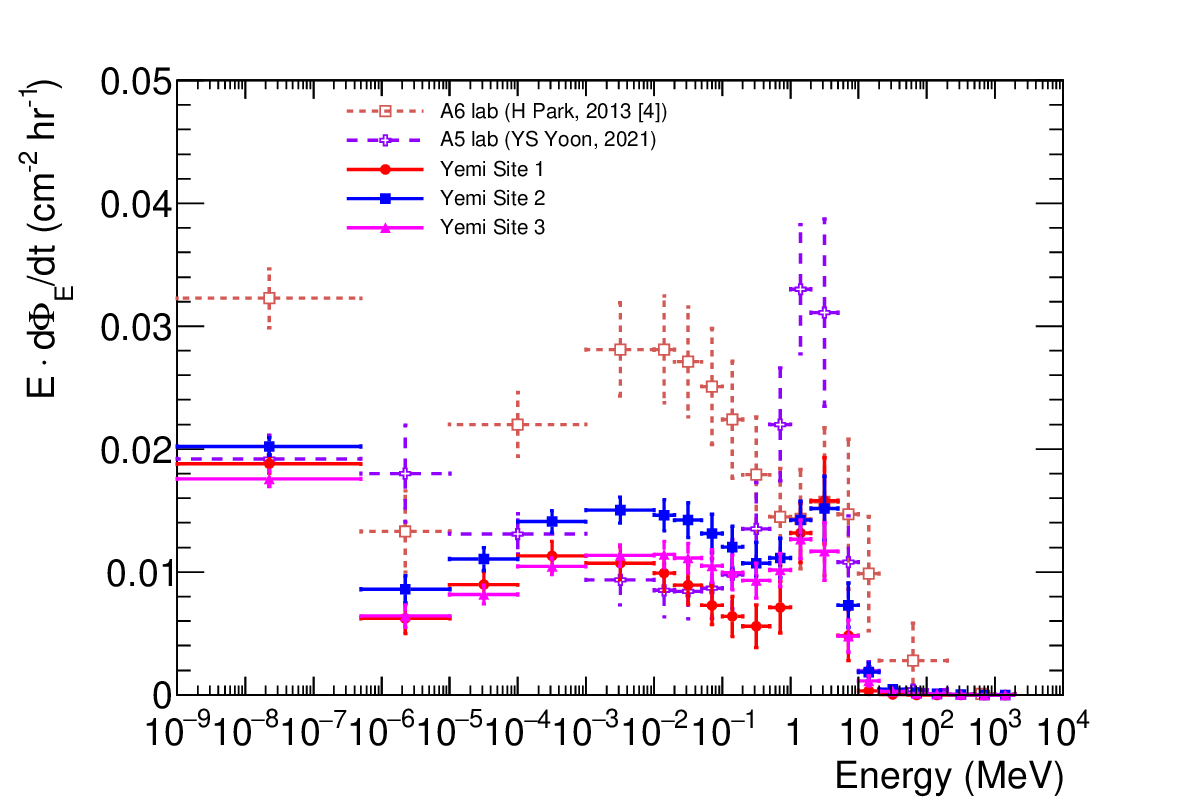}
\caption{Comparison of measured neutron fluence spectra at the Yemilab (filled symbols) and at YangYang Underground Laboratory A5/A6 tunnels (open symbols)~\cite{PARK2013302,Yoon:2021tkv}. The Yemilab results exhibits lower overall neutron background, particularly in the epithermal energy region.}
\label{fig:nspectrum}
\end{figure}

Figure~\ref{fig:nspectrum} compares Yemilab spectra with prior results from the YangYang Underground Laboratory tunnels A5 and A6~\cite{PARK2013302,Yoon:2021tkv}. Neutron fluences in thermal and fast regions at the Yemilab are comparable to those at A6 and A5, respectively. The epithermal region at the Yemilab resembles A6’s spectrum but at lower absolute fluences, indicating an overall reduced neutron background at the Yemilab.

\section{Discussion}

\subsection{Elevated epithermal neutron fluence at Site 2}

The neutron fluence rates measured at the three locations of the Yemilab show differences in both the total and epithermal components. 
Among the sites, Site~2 exhibited a higher total and epithermal neutron fluence compared with Sites~1 and~3. 
This observation is supported by the temporal monitoring data shown in Figure \ref{fig:ratetime}. 
Specifically, detector CKB030, which is particularly sensitive to epithermal neutrons, recorded a distinctively higher event rate at Site 2 compared to the other locations. This excess aligns with the enhanced epithermal component observed in the unfolded spectrum.
Possible explanations for these differences include differences in shotcrete thickness and composition, 
as well as seasonal environmental conditions at the time of measurement.

Site~2 features a shotcrete layer several centimeters thicker than those at the other measurement locations, as reported in private communications with the construction manager~\cite{ParkPrivateComm}. 
The uranium and thorium concentrations in this shotcrete are higher than those typically found in surrounding rock ~\cite{Park:2024sio}. 
The neutron yield via (\alphasym,~n) reactions from uranium decay chains tends to surpass that from thorium for certain materials, such as ceramics and quartz~\cite{KUDRYAVTSEV2020164095,10.1088/1361-6471/adeffa}, so the presence of thicker, U/Th-rich shotcrete likely leads to locally elevated neutron production.

Environmental humidity is another consideration. Measurements at Site~2 were performed in July and August, when relative humidity in the laboratory is higher. 
Simulations of soil moisture effects~\cite{Cirillo2021} and empirical studies in soil measurements~\cite{Schron2017} have demonstrated that increased water content enhances neutron moderation, which boosts the epithermal component at the expense of fast neutrons. 
Thus, the higher humidity at Site~2 may have contributed to the observed enhancement in epithermal neutron rates.

A third possible factor is laboratory activity. Site~2 was undergoing setup for future experiments during the measurement period, involving frequent personnel access and the storage of materials in part of the hall. By contrast, Sites~1 and~3 had limited human traffic. While the effect of such activity on the neutron background is difficult to quantify precisely, it remains a possible source of variation.

In summary, the elevated epithermal neutron flux observed at Site~2 likely arises from a combination of increased neutron production in U/Th-rich shotcrete, enhanced moderation due to seasonal humidity, and possible influences from laboratory activity. Further simulations and sustained monitoring would be required to quantify the contributions of each factor.

\subsection{Comparison with Other Measurements}

Recent studies provide validation and context for our measurement approach and interpretation. 
Plaza et al.~\cite{PLAZA2023102793} investigated thermal neutron backgrounds at the Canfranc Underground Laboratory using both bare and cadmium-shielded \He[3] proportional counters. Their results confirm the utility of waveform-based event selection for effectively separating noise, alpha-induced signals, and neutron events—a method also employed in our work. Moreover, they showed that detectors with stainless-steel bodies exhibit lower internal alpha backgrounds than those with aluminum bodies, supporting our detector choice and background model.
In our case, the internal alpha effects was already reported in earlier measurements at the YangYang Underground Laboratory~\cite{Yoon:2021tkv}, which motivated the dedicated internal alpha background study of the \He[3]-tube detectors described in Section~\ref{sec:internal_alpha}.

\begin{figure} [!htb]
\begin{center}
\includegraphics[width=0.9\textwidth, trim=0 0 0 5,clip ]{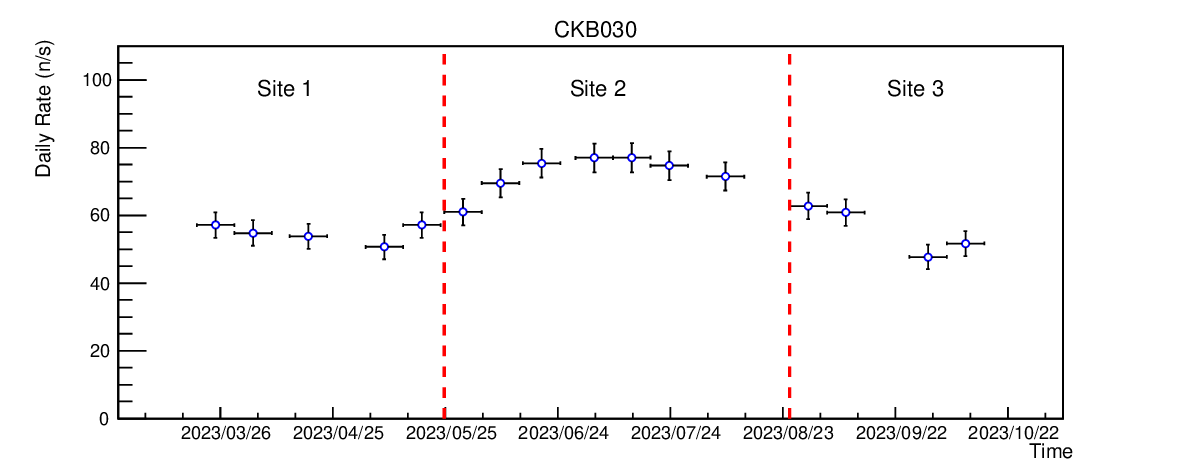}\\
\includegraphics[width=0.9\textwidth, trim=0 0 0 5,clip ]{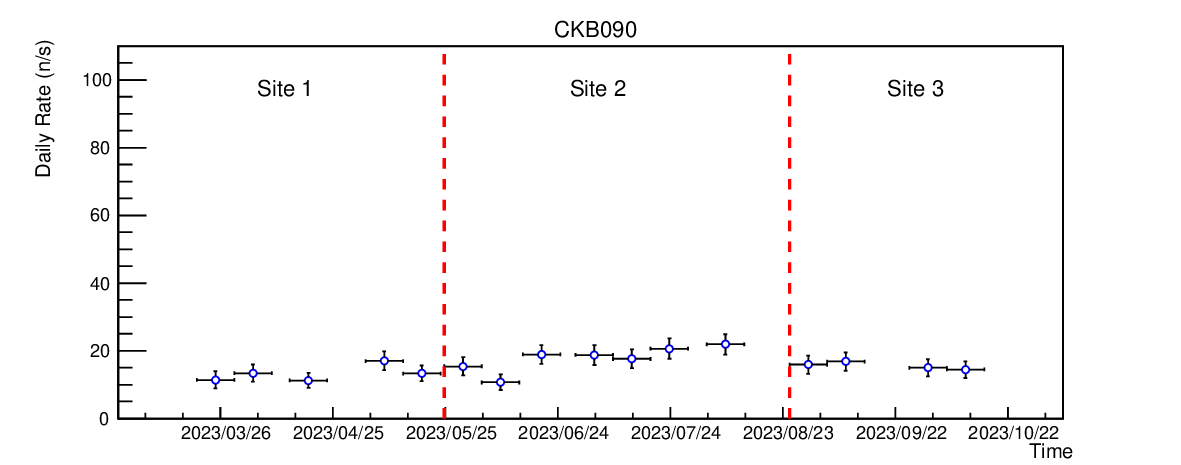}
\caption{Ten-day averaged neutron event rates for detectors CKB030 (top) and CKB090 (bottom) over the measurement period. 
CKB030 shows temporal variation, in contrast to CKB090, which exhibits stable rates.}
\label{fig:ratetime}
\end{center}
\end{figure}

Long-term monitoring at Canfranc by Orrigo et al.~\cite{Orrigo:2022hes} reported that seasonal changes, such as shifts in humidity and radon concentration, drive variations in measured neutron rates. These findings reinforce the importance of seasonal monitoring and align with our observation of time-dependent count rate fluctuations, especially in some detectors at the Yemilab (see Figure~\ref{fig:ratetime}). Although our measurements at the Yemilab were performed sequentially across three sites, the 10-day averaged event rates in select detectors indicate similar temporal variations, as shown in Figure~\ref{fig:ratetime}.

\subsection{Comparison with Other Underground Laboratories}

Table~\ref{tab:fluence} presents neutron fluence rates measured at the Yemilab in the context of data from other major underground laboratories, including LNGS, Modane, Jinping, and Canfranc ~\cite{Belli:1989wz,WULANDARI2004313,Eitel_2012,HU201737,Mei:2009qg,JORDAN20131}. 
The Yemilab’s total neutron fluence rates, $(3.2-4.0) \times 10^{\minus5}\,\mathrm{cm^{\minus2}\,s^{\minus1}}$, are higher than those at deeper sites such as LNGS or Modane, yet lower than those previously obtained at YangYang.
However, observed neutron fluxes do not strictly correlate with overburden depth. At $> 2000~m.w.e.$, the muon-induced component is sub-dominant; instead, the ambient neutron field is primarily driven by (\alphasym{}, n) reactions and spontaneous fission from natural radioactivity (U and Th) in the surrounding rock and tunnel lining materials. This dominance of local radiogenic production explains why facilities with similar depths exhibit varying neutron backgrounds, as the flux depends more on environmental radiopurity than on physical overburden.
 
\begin{table} [!htb]
\begin{center}
\caption{Neutron fluence rates ($\times10^{\minus6}\,\mathrm{cm^{\minus2}\,s^{\minus1}}$) measured using \He[3] proportional counters with moderators at various underground laboratories. Thermal, fast, and total components are listed for comparison; depths are given in meter-water-equivalent (m.w.e.).}
\label{tab:fluence}
\begin{threeparttable}

\begin{tabular}{ccccc}
    \hline
     Underground &  Depth & Thermal   & Fast neutron & Total  \\ 
     laboratory  & (m.w.e.) &  neutron &  [1 - 10 MeV] & neutron  \\\hline \hline
     YangYang A6  \cite{PARK2013302} &  2000 & 24.2 $\pm$ 1.8  & 4.2 $\pm$ 0.9 & 67.2 $\pm$ 2.2  \\
     YangYang A5  \cite{Yoon:2021tkv} &  2000 & 14.4 $\pm$ 1.5  & 7.1 $\pm$ 1.0 & 44.6 $\pm$ 6.6 \\
     Canfranc  \cite{JORDAN20131}  &  2450 & $1.28\pm0.04$  & $1.84\pm0.03$ & $13.8\pm1.4$  \\
     Yemilab Site 1  & 2500 &   14.1 \pmsym\ 0.6  & 3.2 \pmsym\ 0.5 & 32.4 \pmsym\ 1.1 \\
     Yemilab Site 2  & 2500 &  15.1 \pmsym\ 0.5  & 3.4 \pmsym\ 0.4 & 40.1 \pmsym\ 1.0 \\
     Yemilab Site 3  & 2500 &  13.2 \pmsym\ 0.5  & 2.7 \pmsym\ 0.3 & 32.5 \pmsym\ 0.8 \\
     Kamioka  \cite{10.1093/ptep/pty133}   &  2700 & $7.88 $  & $3.88$  & $23.5\pm0.7{}^{+1.9}_{-2.1}$   \\
     LNGS Hall A \cite{Belli:1989wz,WULANDARI2004313} &  3800 & 1.08 $\pm$ 0.02 & 0.7 $\pm$ 0.14 & 3.78 $\pm$ 0.25 \\
     Modane \cite{Eitel_2012}   &  4800 &  3.57$\pm$0.05$\pm$0.27 &  $1.06\pm0.1\pm0.6$ &  9.6  \\
     & & & [ $>$ 1 MeV ] & \\
     CJPL-I \cite{HU201737}    &  6720 & $7.03\pm1.81$  & $3.63\pm2.77$ & $26.9\pm10.2$  \\
     & & & [ 1 - 20 MeV] & \\
    \hline
  \end{tabular}
\end{threeparttable}
\end{center}
\end{table}

\section{Summary}

Neutron fluence was measured at three locations in the Yemilab using a high-sensitivity spectrometer consisting of \He[3] proportional counters and HDPE moderators. The dataset covers a total livetime of 196 days between March and October 2023. The reconstructed neutron energy spectra at all sites and fluences of thermal, epithermal, and fast neutron components were estimated.
Total neutron fluence rates range from $(3.24 \pm 0.11)$ to $(4.01 \pm 0.10) \times 10^{\minus5}~\mathrm{cm^{\minus2}\,s^{\minus1}}$, with Site~2 yielding values approximately 25\% higher than Sites~1 and~3, primarily attributable to increased epithermal neutron rates. Relative to other facilities, the neutron fluence levels at the Yemilab are modestly higher than at LNGS or Modane and lower than those at the YangYang Underground Laboratory.
The neutron background measurements presented here establish baseline conditions at the Yemilab, providing essential parameters for future experimental planning, radiation shielding design, and detector sensitivity optimization—particularly for rare-event physics including dark matter searches and neutrinoless double beta decay studies. Continued neutron monitoring and material assessment will further refine the Yemilab background model and support next-generation physics experiments.

\section*{Acknowledgement}
This work was supported in part by Korea Research Institute of Standards
and Science (KRISS) under Grant GP2025-0008-01 and in part by the National
Research Foundation of Korea funded by the Ministry of Science and ICT
(MSIT) under Grant NRF-2021-R1A2C1-094369.
We acknowledge the support received from the Center for Underground Physics, Institute for Basic Science in Korea.


\bibliography{references}

@article{PARK2013302,
author = "Hyeonseo Park and Jungho Kim and Y.M. Hwang and Kil-Oung Choi",
title = "{Neutron spectrum at the underground laboratory for the ultra low background experiment}",
journal = "Appl. Radiat. Isotopes",
volume = "81",
pages = "302-306",
year = "2013",
issn = "0969-8043",
doi = "10.1016/j.apradiso.2013.03.068"
}

@article{Belli:1989wz,
      author         = "Belli, P. and others",
      title          = "{Deep Underground Neutron Flux Measurement With Large BF-3 Counters}",
      journal        = "Nuovo Cim. A",
      volume         = "101",
      year           = "1989",
      pages          = "959-966",
      doi            = "10.1007/BF02800162"
}

@article{JORDAN20131,
    author = "Jordan, D. and others",
    title = "{Measurement of the neutron background at the Canfranc Underground Laboratory LSC}",
    doi = "10.1016/j.astropartphys.2012.11.007",
    journal = "Astropart. Phys.",
    volume = "42",
    pages = "1--6",
    year = "2013",
    note = "[Erratum: Astropart.Phys. 118, 102372 (2020)]"
}

@article{HU201737,
title = "Neutron background measurements at China Jinping underground laboratory with a Bonner multi-sphere spectrometer",
journal = "Nucl. Instrum. Meth. A",
volume = "859",
pages = "37 - 40",
year = "2017",
issn = "0168-9002",
doi = "10.1016/j.nima.2017.03.048",
author = "Qingdong Hu and others",
}

@article{WULANDARI2004313,
title = "Neutron flux at the Gran Sasso underground laboratory revisited",
      journal        = "Astropart. Phys.",
volume = "22",
number = "3",
pages = "313 - 322",
year = "2004",
issn = "0927-6505",
doi = "10.1016/j.astropartphys.2004.07.005",
author = "H. Wulandari and J. Jochum and W. Rau and F. von Feilitzsch"
}

@article{10.1093/ptep/pty133,
    author = {Mizukoshi, Keita and others},
    title = "{Measurement of ambient neutrons in an underground laboratory at the Kamioka Observatory}",
    journal = {Prog. Theor. Exp. Phys.},
    volume = {2018},
    number = {12},
    year = {2018},
    month = {12},
    issn = {2050-3911},
    doi = {10.1093/ptep/pty133}
}

@article{REGINATTO2002242,
title = "Spectrum unfolding, sensitivity analysis and propagation of uncertainties with the maximum entropy deconvolution code MAXED",
      journal	     = "Nucl. Instrum. Meth. A",
volume = "476",
number = "1",
pages = "242 - 246",
year = "2002",
issn = "0168-9002",
doi = "10.1016/S0168-9002(01)01439-5",
url = "http://www.sciencedirect.com/science/article/pii/S0168900201014395",
author = "Marcel Reginatto and Paul Goldhagen and Sonja Neumann"
}

@article{Eitel_2012,
	doi = {10.1088/1742-6596/375/1/012016},
	url = {https://doi.org/10.1088/1742-6596/375/1/012016},
	year = 2012,
	month = {jul},
	publisher = {{IOP} Publishing},
	volume = {375},
	number = {1},
	pages = {012016},
	author = {Klaus Eitel},
	title = {Measurements of neutron fluxes in the {LSM} underground laboratory},
	journal = {J. Phys. Conf. Ser.}
}

@misc{Centronic,
  author = {{Centronic}},
  title = {Centronic},
  howpublished = {\url{http://www.centronic.co.uk/}},
  year = {2020},
  note = {Accessed: 2020-07-15}
}

@article{WIEGEL200236,
title = "NEMUS—the PTB Neutron Multisphere Spectrometer: Bonner spheres and more",
journal = "Nuclear Instruments and Methods in Physics Research Section A: Accelerators, Spectrometers, Detectors and Associated Equipment",
volume = "476",
number = "1",
pages = "36 - 41",
year = "2002",
issn = "0168-9002",
doi = "https://doi.org/10.1016/S0168-9002(01)01385-7",
url = "http://www.sciencedirect.com/science/article/pii/S0168900201013857",
author = "B Wiegel and A.V Alevra"
}

@article{Park:2024sio,
    author = "Park, K. S. and Kim, Y. D. and Bang, K. M. and Park, H. K. and Lee, M. H. and So, J. and Kim, S. H. and Jang, J. H. and Kim, J. H. and Kim, S. B.",
    title = "{Construction of Yemilab}",
    eprint = "2402.13708",
    archivePrefix = "arXiv",
    primaryClass = "astro-ph.IM",
    doi = "10.3389/fphy.2024.1323991",
    journal = "Front. in Phys.",
    volume = "12",
    pages = "1323991",
    year = "2024"
}

@article{Yoon:2021tkv,
    author = "Yoon, Young Soo and Kim, Jungho and Park, Hyeonseo",
    title = "{Neutron background measurement for rare event search experiments in the YangYang underground laboratory}",
    eprint = "2102.07205",
    archivePrefix = "arXiv",
    primaryClass = "nucl-ex",
    doi = "10.1016/j.astropartphys.2020.102533",
    journal = "Astropart. Phys.",
    volume = "126",
    pages = "102533",
    year = "2021"
}

@article{Mei:2009qg,
  author       = {Mei, D.-M. and Hime, A.},
  title        = {Muon-induced background study for underground laboratories},
  journal      = {Phys. Rev. D},
  volume       = {73},
  pages        = {053004},
  year         = {2006},
  doi          = {10.1103/PhysRevD.73.053004}
}

@article{Orrigo:2022hes,
    author = "Orrigo, S. E. A. and others",
    title = "{Long-term evolution of the neutron rate at the Canfranc Underground Laboratory}",
    eprint = "2204.14263",
    archivePrefix = "arXiv",
    primaryClass = "nucl-ex",
    doi = "10.1140/epjc/s10052-022-10755-6",
    journal = "Eur. Phys. J. C",
    volume = "82",
    number = "9",
    pages = "814",
    year = "2022"
}

@article{Cirillo2021,
  author    = {Andrea Cirillo and Ruggero Meucci and Michele Caresana and Marco Caresana},
  title     = {An innovative neutron spectrometer for soil moisture measurements},
  journal   = {Eur. Phys. J. Plus},
  volume    = {136},
  number    = {985},
  year      = {2021},
  doi       = {10.1140/epjp/s13360-021-01976-x},
  url       = {https://doi.org/10.1140/epjp/s13360-021-01976-x}
}

@article{KUDRYAVTSEV2020164095,
title = {Neutron production in (α,n) reactions},
journal = {Nuclear Instruments and Methods in Physics Research Section A: Accelerators, Spectrometers, Detectors and Associated Equipment},
volume = {972},
pages = {164095},
year = {2020},
issn = {0168-9002},
doi = {https://doi.org/10.1016/j.nima.2020.164095},
url = {https://www.sciencedirect.com/science/article/pii/S0168900220305088},
author = {V.A. Kudryavtsev and P. Zakhary and B. Easeman}
}

@article{10.1088/1361-6471/adeffa,
      author={Santorelli, Roberto and Cano-Ott, Daniel and Cebrian, S and Dimitriou, Paraskevi and Gromov, Maxim and Harańczyk, Małgorzata and Kish, Alex and Kluck, Holger and Kudryavtsev, Vitaly A and Lazanu, Ionel and Lozza, Valentina and Luzón Marco, Gloria and Mendoza Cembranos, Emilio and Parvu, Mihaela and Pesudo, Vicente and Pocar, Andrea and Selvi, Marco and Westerdale, Shawn and Zuzel, Grzegorz},
      title={Review of Neutron Yield from (α, n) Reactions: Data, Methods, and Prospects},
      journal={Journal of Physics G: Nuclear and Particle Physics},
      url={http://iopscience.iop.org/article/10.1088/1361-6471/adeffa},
      year={2025}
}

@article{PLAZA2023102793,
title = {Thermal neutron background at Laboratorio Subterráneo de Canfranc (LSC)},
journal = {Astroparticle Physics},
volume = {146},
pages = {102793},
year = {2023},
issn = {0927-6505},
doi = {https://doi.org/10.1016/j.astropartphys.2022.102793},
url = {https://www.sciencedirect.com/science/article/pii/S0927650522000949},
author = {J. Plaza and T. Martínez and V. Bécares and D. Cano-Ott and D. Villamarín and A. {Pérez de Rada} and E. Mendoza and V. Pesudo and R. Santorelli and C. Peña and J. Balibrea-Correa and A. Boeltzig}
}

@article{Schron2017,
  author    = {Andreas Schrön and Mareike Köhli and Markus Schröter and Matthias S. Zreda and Sascha Schmidt and Sascha Dietrich and Martijn W. Groenendijk and Markus J. Hendricks Franssen},
  title     = {Cosmic-ray neutron sensing of soil moisture: a review of the state of the art},
  journal   = {Hydrology and Earth System Sciences},
  volume    = {21},
  pages     = {5009--5034},
  year      = {2017},
  doi       = {10.5194/hess-21-5009-2017}
}

@misc{ParkPrivateComm,
  author       = {K.S. Park},
  title        = {Private communication},
  year         = {2024},
  note         = {Information regarding thicker shotcrete layers in the AMoRE Hall at the Yemi Underground Laboratory}
}

@ARTICLE{2024arXiv240913226C,
       author = {{Carlin}, N. and {Cho}, J.~Y. and {Choi}, J.~J. and {Choi}, S. and {Ezeribe}, A.~C. and {Franca}, L.~E. and {Ha}, C. and {Hahn}, I.~S. and {Hollick}, S.~J. and {Jeon}, E.~J. and {Joo}, H.~W. and {Kang}, W.~G. and {Kauer}, M. and {Kim}, B.~H. and {Kim}, H.~J. and {Kim}, J. and {Kim}, K.~W. and {Kim}, S.~H. and {Kim}, S.~K. and {Kim}, W.~K. and {Kim}, Y.~D. and {Kim}, Y.~H. and {Ko}, Y.~J. and {Lee}, D.~H. and {Lee}, E.~K. and {Lee}, H. and {Lee}, H.~S. and {Lee}, H.~Y. and {Lee}, I.~S. and {Lee}, J. and {Lee}, J.~Y. and {Lee}, M.~H. and {Lee}, S.~H. and {Lee}, S.~M. and {Lee}, Y.~J. and {Leonard}, D.~S. and {Luan}, N.~T. and {Machado}, V.~H.~A. and {Manzato}, B.~B. and {Maruyama}, R.~H. and {Neal}, R.~J. and {Olsen}, S.~L. and {Park}, B.~J. and {Park}, H.~K. and {Park}, H.~S. and {Park}, J.~C. and {Park}, K.~S. and {Park}, S.~D. and {Pitta}, R.~L.~C. and {Prihtiadi}, H. and {Ra}, S.~J. and {Rott}, C. and {Shin}, K.~A. and {Cavalcante}, D.~F.~F.~S. and {Son}, M.~K. and {Spooner}, N.~J.~C. and {Truc}, L.~T. and {Yang}, L. and {Yu}, G.~H.},
        title = "{COSINE-100 Full Dataset Challenges the Annual Modulation Signal of DAMA/LIBRA}",
      journal = {arXiv e-prints},
         year = 2024,
        month = sep,
          eid = {arXiv:2409.13226},
        pages = {arXiv:2409.13226},
          doi = {10.48550/arXiv.2409.13226},
archivePrefix = {arXiv},
       eprint = {2409.13226},
 primaryClass = {hep-ex},
       adsurl = {https://ui.adsabs.harvard.edu/abs/2024arXiv240913226C}
}

@article{PhysRevLett.134.082501,
  title = {Improved Limit on Neutrinoless Double Beta Decay of $^{100}\mathrm{Mo}$ from AMoRE-I},
  author = {Agrawal, A. and Alenkov, V. V. and Aryal, P. and Beyer, J. and Bhandari, B. and Boiko, R. S. and Boonin, K. and Buzanov, O. and Byeon, C. R. and Chanthima, N. and Cheoun, M. K. and Choe, J. S. and Choi, Seonho and Choudhury, S. and Chung, J. S. and Danevich, F. A. and Djamal, M. and Drung, D. and Enss, C. and Fleischmann, A. and Gangapshev, A. M. and Gastaldo, L. and Gavrilyuk, Y. M. and Gezhaev, A. M. and Gileva, O. and Grigorieva, V. D. and Gurentsov, V. I. and Ha, C. and Ha, D. H. and Ha, E. J. and Hwang, D. H. and Jeon, E. J. and Jeon, J. A. and Jo, H. S. and Kaewkhao, J. and Kang, C. S. and Kang, W. G. and Kazalov, V. V. and Kempf, S. and Khan, A. and Khan, S. and Kim, D. Y. and Kim, G. W. and Kim, H. B. and Kim, Ho-Jong and Kim, H. J. and Kim, H. L. and Kim, H. S. and Kim, M. B. and Kim, S. C. and Kim, S. K. and Kim, S. R. and Kim, W. T. and Kim, Y. D. and Kim, Y. H. and Kirdsiri, K. and Ko, Y. J. and Kobychev, V. V. and Kornoukhov, V. and Kuzminov, V. V. and Kwon, D. H. and Lee, C. H. and Lee, DongYeup and Lee, E. K. and Lee, H. J. and Lee, H. S. and Lee, J. and Lee, J. Y. and Lee, K. B. and Lee, M. H. and Lee, M. K. and Lee, S. W. and Lee, Y. C. and Leonard, D. S. and Lim, H. S. and Mailyan, B. and Makarov, E. P. and Nyanda, P. and Oh, Y. and Olsen, S. L. and Panasenko, S. I. and Park, H. K. and Park, H. S. and Park, K. S. and Park, S. Y. and Polischuk, O. G. and Prihtiadi, H. and Ra, S. and Ratkevich, S. S. and Rooh, G. and Sala, E. and Sari, M. B. and Seo, J. and Seo, K. M. and Sharma, B. and Shin, K. A. and Shlegel, V. N. and Siyeon, K. and So, J. and Sokur, N. V. and Son, J. K. and Song, J. W. and Srisittipokakun, N. and Tretyak, V. I. and Wirawan, R. and Woo, K. R. and Yeon, H. J. and Yoon, Y. S. and Yue, Q.},
  collaboration = {AMoRE Collaboration},
  journal = {Phys. Rev. Lett.},
  volume = {134},
  issue = {8},
  pages = {082501},
  numpages = {8},
  year = {2025},
  month = {Feb},
  publisher = {American Physical Society},
  doi = {10.1103/PhysRevLett.134.082501},
  url = {https://link.aps.org/doi/10.1103/PhysRevLett.134.082501}
}

\end{document}